# Cylindrical quantum well of finite depth in external magnetic field.

Lobanova O.R.[a,1], Ivanov A.I.[b]

[a,b]Kaliningrad State University, Theoretical Physics Department, Kaliningrad, Russia.

**Abstract**

Energy spectrum of an electron confined by finite hard-wall potential in a cylinder quantum dot placed in weak (up to 100 kOe) homogeneous external magnetic field were calculated using the method of variation of vector potential. Electronic motion along the cylinder axis is limited by one-dimensional infinite potential barrier and electronic motion on the plane perpendicular to the axis is limited by two-dimensional finite potential barrier.

## 1.Introduction

Recent progress in nanotechnology allows creating semiconductor nanostructures – quantum dots (QDs), wells or wires [1-3] - where electrons are confined in one, two or all three directions. The electronic properties of dots show many parallels with those of atoms. Most notably, the confinement of the electrons in all three spatial directions results in a quantized energy spectrum [4-6]. Quantum dots are therefore regarded as artificial atoms. Such nanostructures are suitable for experiments that cannot be carried out in atomic physics, in particular due to the high level of control. This has given rise to the possibility of a number of very interesting applications, such as producing single-electron transistors [7], quantum dot lasers [8], etc. Spherelike, circular cylinderlike or corral shaped enclosed structures were widely investigated in nanometer scale [9-15]. It is especially interesting to observe the effect of a magnetic field on the atomic-like properties (energy spectrum, magnetic susceptibilities, etc.) [16-19].

---

[1] Corresponding author: tel.: +7(0112)338217; fax: +7(0112)465813. E-mail address: olgalobanova@yahoo.com

Effective method of calculation of weak external magnetic field influence on energy levels in quantum dot is method of variation of vector potential, offered by T.K. Rebane [20]. Earlier this method was successfully applied to calculation of proton nuclear magnetic shielding constants in diatomic molecules [21] and to calculation of energy levels of an electron confined in a closed region of space with cylindrical geometry [16].

In the present work we have investigated the energy spectrum of one electron circular cylinder quantum dot with a hard-wall confinement in the presence of an external magnetic field. Along the cylinder axis an electron is confined by infinite potential barrier, whereas on the plane perpendicular to the axis electron is confined by potential barrier of finite height. In such statement of the problem the Schroedinger equation for circular cylinder QD in cylindrical coordinates is separable as for the extreme case of an infinite potential well. It is necessary to emphasize that for the analogous (circular cylinder) problem with finite barrier in all three directions the Schroedinger equation is not separable as mistakenly asserted in [14]. Applied magnetic field is directed perpendicular to the cylinder axis and considered to be weak (up to 100 kOe), fixed and homogeneous.

## 2. Calculation method and results

We consider an electron confined in circular cylinder of radius R and height l. The electron interacts with a potential:

$$V = \begin{cases} V_0, \rho > R, z \in (0;l), \\ 0, \rho \in (0;R), z \in (0;l), \\ \infty, z < 0, z > l. \end{cases} \quad (1)$$

For given cylindrical QD the separation of variables of Schroedinger equation in cylindrical coordinates leads to two equations: one equation contains z-variable and can easily be solved; other equation in XY-plane contains angular and radial parts. Solutions of radial part are Bessel functions. Energy spectrum in XY-plane $E_{xy}$ is obtained from the continuity condition of logarithmic derivative at the boundary of the well ($\rho = R$):

$$\frac{J'_m(kR)}{J_m(kR)} = \frac{K'_m(\aleph R)}{K_m(\aleph R)}, \quad (2)$$

where $J_m(kR), K_m(\aleph R)$ are Bessel functions, $k = \sqrt{2E_{xy}}, \aleph = \sqrt{2(V_0 - E_{xy})}$ (in a.u.). Corresponding eigen-equations are

$$\begin{aligned} m = 0 &: kK_0(\aleph R)J_1(kR) = \aleph J_0(kR)K_1(\aleph R), \\ m > 0 &: kK_m(\aleph R)J_{m-1}(kR) = \aleph J_m(kR)K_{m-1}(\aleph R). \end{aligned} \quad (3)$$

Total energy $E = E_{xy} + E_z$.

Further given artificial atom is placed in external magnetic field $\vec{H}$ directed perpendicular to the cylinder axis with vector potential $\vec{A}$. Vector potential $\vec{A}$ satisfies the condition $rot\, \vec{A} = \vec{H}$. This condition defines vector potential to within the gradient of arbitrary function of electron coordinates. Hence all vector potentials of the class

$$\vec{A} = \frac{1}{2}[\vec{H} \times \vec{r}] + \vec{\nabla} f(\vec{r}), \tag{4}$$

where $f(\vec{r})$ - arbitrary doubly differentiable function, physically equal: they describe the same magnetic field $\vec{H}$. Because of gradient invariance of Schroedinger equation electron energy and corrections to energy are invariant relatively the choice of function $f(\vec{r})$.

It is well known that in case of weak magnetic field corrections to energy in the first and second orders of perturbation theory are as follows:

$$\begin{aligned} E^{(H)} &= \int \Psi^{(0)*} W^H \Psi^{(0)} dV, \\ E^{(H^2)} &= E_d^{(H^2)} + E_p^{(H^2)}, \\ E_d^{(H^2)} &= \int \Psi^{(0)*} W^{H^2} \Psi^{(0)} dV, \quad E_p^{(H^2)} = \int \Psi^{(0)*} W^H \Psi^{(H)} dV, \end{aligned} \tag{5}$$

where $E_d^{(H^2)}$, $E_p^{(H^2)}$ - diamagnetic and paramagnetic corrections to energy respectively, $W^H$, $W^{H^2}$ - linear and quadratic relatively magnetic field intensity $\vec{H}$ perturbation operators (in a.u.)

$$W^H = -\frac{i}{2c}(2\vec{A}\nabla + \text{div}\vec{A}), \quad W^{H^2} = \frac{1}{2c^2}\vec{A}^2. \tag{6}$$

In given problem unperturbed wave function $\Psi^{(0)}$ is real, hence value $E^{(H)}$ turns to zero ($W^H$ is imaginary). To calculate quadratic correction $E^{(H^2)}$ it is necessary to know correction wave function $\Psi^{(H)}$, which can be obtained even approximately only in simplest cases.

In method of variation of vector potential [20, 21] calculation of correction to wave function in magnetic field $\Psi^{(H)}$ is replaced by calculation of optimum function of gradient transformation $f_{opt}$ within the variation procedure. The optimum function $f_{opt}$ is consistent with optimum vector potential $\vec{A}_{opt} = \frac{1}{2}[\vec{H} \times \vec{r}] + \vec{\nabla} f_{opt}$. Within this method calculation of corrections to energy, caused by external magnetic field, is reduced to calculation of corresponding diamagnetic quantities with $\vec{A}_{opt}$.

Diamagnetic correction to energy, corresponding to vector potential (2) (in a.u.) is

$$J(f) = \frac{1}{2c^2}\int |\Psi^{(0)}|^2 (\frac{1}{2}[\vec{H} \times \vec{r}] + \vec{\nabla} f)^2 dV. \tag{7}$$

Value $J(f)$ is a functional relative to function of gradient transformation $f$. From stationarity condition of functional $J(f)$

$$\delta J(f) = 0 \tag{8}$$

it is possible to obtain optimum function of gradient transformation $f_{opt}$. Minimum value of functional $J(f)$ equals to total quadratic relatively magnetic field intensity correction to energy

$$E^{(H^2)} = \min J(f). \tag{9}$$

Quadratic correction to energy $E^{(H^2)}$ is completely determined by unperturbed wave function $\Psi^{(0)}$.

Minimum of functional $J(f)$ usually is defined approximately, varying function of gradient transformation $f$ in some limited class of functions.

Energy levels of electron in cylinder QD were calculated in circular and elliptic approximations. Magnetic field was directed along Y-axis: $\vec{H} = (0, H, 0)$.

In *circular approximation* one may find (in a.u.)

$$\vec{\nabla} f_{1opt} = -H\bar{z}\vec{i}, \tag{10}$$

where $\bar{z} = \int |\Psi^{(0)}|^2 z \, dV$.

Therefore quadratic relatively magnetic field intensity correction to energy in circular approximation has a form

$$E_1^{(H^2)} = \frac{H^2}{2c^2}\left\{\overline{x^2} + \overline{z^2} - \bar{z}^2\right\}, \tag{11}$$

where $\overline{x^2} = \int |\Psi^{(0)}|^2 x^2 \, dV$, $\overline{z^2} = \int |\Psi^{(0)}|^2 z^2 \, dV$.

In *elliptic approximation* we have (in a.u.)

$$\vec{\nabla} f_{2opt} = H\left\{-\frac{\overline{x^2}\,\bar{z}}{\overline{x^2}+\overline{z^2}-\bar{z}^2} + \frac{\overline{x^2}-(\overline{z^2}-\bar{z}^2)}{\overline{x^2}+\overline{z^2}-\bar{z}^2}z\right\}\vec{i} + H\frac{\overline{x^2}-(\overline{z^2}-\bar{z}^2)}{\overline{x^2}+\overline{z^2}-\bar{z}^2}x\vec{k}. \tag{12}$$

Thus in elliptic approximation correction to energy is

$$E_2^{(H^2)} = \frac{H^2}{2c^2}\left\{\frac{\overline{x^2}\,\bar{z}^2}{(\overline{x^2}+\overline{z^2}-\bar{z}^2)^2} + 4\frac{\overline{z^2}-\bar{z}^2}{\overline{x^2}+\overline{z^2}-\bar{z}^2}z\right\}. \tag{13}$$

Electron energy spectrum in the presence of magnetic field is shown in Table 1. Every energy level is characterized by three quantum numbers: $m = 0, 1, 2..$; $k=1, 2,..$ (numbers zeros of Bessel functions); $k' = 1, 2,..$ (numbers zeros of sine). It is convenient to set obtained energy levels in ascending order of their values. In present paper first 10 low-lying energy levels are shown.

Parameters, used in these calculations, are $R = 2.75$ nm; $l = 4$ nm; $V_0 = 1$ eV ; $H = 100$ kOe.

**Table 1**: Quantum dot energy levels (in meV) in magnetic field

| n | $E_{0,n}$ | $E_{10,n}$ | $E_{20,n}$ | $E_{1,n}$ | $E_{11,n}$ | $E_{21,n}$ | $E_{2,n}$ | $E_{12,n}$ | $E_{22,n}$ |
|---|---|---|---|---|---|---|---|---|---|
| 1 | 57 | 59 | 60 | 109 | 111 | 112 | 178 | 180 | 181 |
| 2 | 128 | 130 | 129 | 180 | 182 | 182 | 249 | 251 | 251 |
| 3 | 202 | 204 | 205 | 297 | 300 | 299 | 366 | 369 | 368 |
| 4 | 245 | 247 | 247 | 313 | 315 | 316 | 442 | 444 | 445 |
| 5 | 273 | 275 | 275 | 384 | 386 | 386 | 513 | 515 | 515 |
| 6 | 390 | 393 | 392 | 462 | 464 | 464 | 531 | 533 | 533 |
| 7 | 410 | 412 | 411 | 501 | 504 | 503 | 630 | 633 | 632 |
| 8 | 467 | 469 | 470 | 658 | 660 | 661 | 717 | 719 | 720 |
| 9 | 537 | 540 | 539 | 666 | 668 | 668 | 742 | 745 | 744 |
| 10 | 555 | 557 | 557 | 673 | 676 | 675 | 788 | 790 | 790 |

$E_{m,n}$ - unperturbed energy levels; $E_{1m,n}$, $E_{2m,n}$ - energy levels in magnetic field calculated in circular and elliptic approximations respectively (*n* is not a quantum number, index *n* numbers first 10 low-lying energy levels in ascending order and corresponds to definite combination *{k, k'}*). States with *m* = 0, 1, 2 are considered.

## 3. Conclusions

Calculation of energy spectrum of one electron circular cylinder quantum dot in weak homogeneous external magnetic field was implemented. Along the cylinder axis an electron was confined by infinite potential barrier, whereas on the plane perpendicular to the axis electron was confined by potential barrier of finite height.

In the first order of perturbation theory correction to energy linear relatively magnetic field intensity turns into zero. Second order of perturbation theory leads to shifts of energy levels in magnetic field. Quadratic relatively magnetic field intensity corrections to energy, obtained in circular and elliptic approximations, weakly differ from each other.

## References

[1] T.Chakraborty, Comments Condens. Matter Phys. **16**, (1992), 35.
[2] R.Turton, The Quantum Dot. A Journey into future microelectronics, Oxford Univ. Press, NY (1995).
[3] L.Jacak, P.Hawrylak, A.Wojs, Quantum Dots, Springer-Verlag, Berlin (1997).
[4] M.Macucci, K.Hess, G.J.Iafrate, Phys.Rev.B **48**, (1993), 17354.
[5] W.D.Heiss, R.G.Nazmitdinov, Phys.Lett.A **222**, (1996), 309; Phys.Rev.B **55**, (1997), 16310.
[6] S.Tarucha, D.G.Austing, T.Honda, Phys.Rev.Lett. **77**, (1996), 3613.
[7] Zhuang Lei, Guo Lingjie, Chou S.Y., Appl.Phys.Lett. **72**, (1998), 1205.
[8] Grundmann M., Physica E **5**, (2000), 167.
[9] M. F. Crommie, C. P. Lutz, and D. M. Eigler, Science **262**, (1993), 218.
[10] H. Masuda and K. Fukuda, Science **268**, (1995), 1466.
[11] D. H. Pearson and R. J. Tonucci, Science **270**, (1995), 68.
[12] M. Tadic, F.M. Peeters, K.L. Janssens, M. Korkusinski, P. Hawrylak, J. Appl. Phys. **92**, (2002), 5819-5829.
[13] M. Tadic, F.M. Peeters, K.L. Janssens, Phys. Rev. B **65**, (2002), 165333.

[14] Yong Fang Zhao, Xiao Gong Jing, Lin Song Li, Li Jun Wang, Zheng Hui, Tie Jin Li, J. Vac. Sci. Technol. B **15**(4), Jul/Aug (1997).
[15] G.Todorovic, V.Milanovic, Z.Ikonic, D.Indjin, Solid State Comm. **103**, №5, (1997), 319-323; J. Vac. Sci. Technol. B **15**(4), Jul/Aug (1997).
[16] A.I.Ivanov, O.R.Lobanova, Physica E **23**, (2004) 61-64.
[17] M.B. Tavernier, E. Anisimiovas, F.M. Peeters, B. Szafran, J. Adamowski, S. Bednarek, Phys. Rev. B **68**, (2003), 205305.
[18] K.L. Janssens, B. Partoens, F.M. Peeters, Phys. Rev. B **67**, (2003), 235325.
[19] F.M. Peeters V.A. Schweigert, Phys. Rev. B **53**, (1996), 1468-1474.
[20] Rebane T.K., JETP **38**, №3, (1960), 963-965.
[21] Rebane T.K., Vestnik Leningradskogo Universiteta, №4, (1973), 38-45.